\documentclass[11pt,english]{article}
\usepackage[utf8]{inputenc}
\usepackage{amsmath}
\usepackage{amsthm}
\usepackage{amssymb}

\makeatletter

\DeclareTextSymbolDefault{\textquotedbl}{T1}
\providecommand{\tabularnewline}{\\}

\@ifundefined{date}{}{\date{}}
\usepackage{t1enc}
\usepackage[english]{babel}
\usepackage{amsthm}
\usepackage{braket}

\def\cros{\raise1.9pt\hbox{$\scriptscriptstyle
          >$}\!\raise1.5pt\hbox{$\scriptstyle\triangleleft\,$}}

\theoremstyle{definition}\theoremstyle{definition}\theoremstyle{definition}\theoremstyle{definition}\newcommand{\noi}{\vspace{0.1in} \noindent}

\title{\bf Sequential measurements \\ and the Kochen-Specker arguments}
\author{\textit{Gábor Hofer-Szabó}\thanks{Research Center for the Humanities, Budapest, email: szabo.gabor@btk.mta.hu}}

\usepackage{babel}

\makeatother

\usepackage{babel}
\begin{document}
\maketitle 
\begin{abstract}
It will be shown that the Peres-Mermin square admits value-definite
noncontextual hidden-variable models if the observables associated
with the operators can be measured only sequentially but not simultaneously.
Namely, sequential measurements allow for noncontextual models in
which hidden states update between consecutive measurements. Two recent
experiments realizing the Peres-Mermin square by sequential measurements
will also be analyzed along with other hidden-variable models accounting
for these experiments.\vspace{0.1in}

\end{abstract}

\section{Introduction}

How can we justify that in every hidden (ontic) state two observables
represented by commuting operators in quantum mechanics have joint
values corresponding to the eigenvalues in one of the common eigenstates
of the operators? Well, there is no other way than to measure the
two observables \emph{simultaneously} in various quantum states, eigenstates
and non-eigenstates, and to check the joint outcomes directly. If
the joint outcomes conform to these eigenvalues, then (assuming that
quantum states are just distributions of hidden states) we can be
pretty sure that in every hidden state the observables have just those
joint values. 

But what if the measurement of the observables can be performed only
\emph{sequentially}, that is only one after the other? In this case
quantum mechanics tells us that the quantum state will update upon
measurements according to the projection postulate and the subsequent
outcomes will again conform to just the eigenvalues, irrespective
of the order of the measurements. But does it mean that the observables
have the joint values corresponding to those eigenvalues in each step
of the measurement process? No, it does not. Generally one cannot
draw a conclusion from diachronic evidences to synchronic facts. And
indeed, as we will shortly see, one can easily cook up hidden-variable
models for commuting operators such that the simultaneous joint values
do not conform to the above eigenvalues, still the subsequent outcomes
of the measurements do. 

Why does this problem matter?

In the Kochen-Specker theorems one proves that there is no value assignment
for certain tricky sets of operators such that the values assigned
to any subset of mutually commuting operators conform to the eigenvalues
in one of the common eigenstates of the operators in the subset. These
value assignments represent hidden states in a value-definite (outcome-deterministic)
hidden-variable model and the values assigned to the commuting operators
represent the joint values of the corresponding observables. Since
the no-go theorems rule out such value assignments, they also rule
out value-definite hidden-variable models.

These models are also called noncontextual. The reason for this is
that the commuting operators are associated with simultaneous measurements
which, if performed, yield just the outcomes corresponding to the
above joint values. The model is noncontextual in the sense that the
value of an observable and the outcome of the associated measurement
is just the value assigned to the operator in the given hidden state,
irrespective of whether a simultaneous measurement is performed or
not (see also Hofer-Szabó, 2021a,b, 2022). In short, the no-go theorems
show that there is no hidden-variable model which would assign values
to observables which are independent of simultaneous measurements,
that is which are noncontextual. 

But what if the measurements realizing commuting operators cannot
be performed simultaneously but only sequentially? As I claimed above
and will show below, in this case the central assumption of the Kochen-Specker
theorems---namely that the joint values assigned to commuting operators
should conform to the eigenvalues in one of the common eigenstates
of the operators---remains physically unjustified. In case of sequential
measurements, the joint values can freely update in such a way that
\emph{the sequential outcomes do but the simultaneous joint values
do not conform to these eigenvalues}. But if the joint values need
not conform to the eigenvalue-constraint, the no-go theorems will
not go through, hence opening the way for a noncontextual value-definite
hidden-variable model. 

Let me be clear already from the outset what kind of noncontextual
value-definite hidden-variable models will be constructed. These models
are value-definite in the sense that in every hidden state each observable
has a definite value which is simply revealed if measured. The model
is noncontextual in the usual sense: the value revealed by a measurement
does not depend on whether other simultaneous measurements are performed
or not. Now, the only fly in the ointment is that in each hidden state
there are observables associated with commuting operators which have
joint values \emph{not} conforming to the eigenvalues in one of the
common eigenstates of the operators. If these observables could be
measured \emph{simultaneously}, this anomaly would be revealed. But
if they can be measured only \emph{sequentially}, then the hidden
state can change upon the measurements such that this anomaly remains
concealed throughout the whole measurement process. 

In some recent experiments devised to verify the Kochen-Specker arguments,
simultaneous measurements are replaced, due to technological reasons,
by sequential measurements. My paper is directed against these experiments.
The central claim of the paper is the following: \emph{if the operators
featuring in a Kochen-Specker theorem are realized by measurements
which can be performed only sequentially but not simultaneously, then
the argument does not rule out noncontextual value-definite hidden-variable
models in which the hidden state can update upon the subsequent measurements.}

This paper intends to contribute to the debate on the experimental
testability of the Kochen-Specker theorems via \emph{sequential} measurements.
The question of empirical testability of the Kochen-Specker theorems
is not new (see for example Held, 2022, Sec. 6 for the various experimental
challenges). What is new, however, is the awareness of a kind of loophole
in the argument provided by the experimental fact that the commuting
operators are realized by sequential measurements. Whereas simultaneous
measurements provide \emph{synchronic} \emph{constrains} on the possible
hidden states which constraints then lead to the Kochen-Specker arguments,
sequential measurements provide only \emph{diachronic} \emph{constrains}:
the outcome statistics of the consecutive measurements need to conform
to the projection postulate. This opens the way to construct hidden-variable
models which avoid the synchronic constraints and satisfy the diachronic
ones. And indeed, some authors (La Cour 2009, 2017) developed highly
sophisticated hidden-variable models for various experimental tests
with sequential measurements. Others (Gühne et al., 2010) derived
generalized Kochen-Specker inequalities for sequential measurements
not strictly subscribing to the projection postulate. Interestingly,
the various models use different concepts of noncontextuality, some
of which are stricter, some are weaker than the one used in this paper.
In the Discussion, I will situate my approach in this wider context
of noncontextual hidden-variable models for sequential tests of the
Kochen-Specker theorems. 

In the paper I will proceed as follows. In Section 2, as a warm-up
exercise, I construct a noncontextual value-definite hidden-variable
model for three commuting operators realized by sequential measurements.
In Section 3, a similar model for the entire Peres-Mermin square will
be constructed. In Section 4, I analyze two recent experiments realizing
the Peres-Mermin argument by sequential measurements. In Section 5,
the various concepts of noncontextuality used in the literature and
the various sequential models will be compared. I conclude in Section
6.

\section{A simple example}

Consider the following three pairwise commuting self-adjoint operators: 
\begin{center}
\begin{tabular}{lll}
\quad{}$A_{1}=\sigma_{x}\otimes\sigma_{x}$ \quad{}  & \quad{}$A_{2}=\sigma_{y}\otimes\sigma_{y}$ \quad{}  & \quad{}$A_{3}=\sigma_{z}\otimes\sigma_{z}$ \quad{}\tabularnewline
\end{tabular}
\par\end{center}

\noindent where $\sigma_{x}$, $\sigma_{y}$, and $\sigma_{z}$ are
the Pauli operators on the two dimensional complex Hilbert space $H_{2}$.
Suppose we realize the operators by three measurements $M_{i}$ ($i=1,2,3)$
with outcomes $O_{i}^{\pm}$. The realization is successful if for
each individual measurement $M_{i}$:

\begin{equation}
\braket{\Psi\vert P_{i}^{\pm}\Psi}=p_{\Psi}(O_{i}^{\pm}\,\vert\,M_{i})\label{eq:QM1}
\end{equation}
where $P_{i}^{\pm}$ denotes the eigenprojections of the operator
$A_{i}$ with eigenvalue $\pm1$ and $p_{\Psi}$ denotes the probability
of outcomes of measurements performed on a system prepared in quantum
state $\ket{\Psi}$. 

First, suppose that $M_{1}$, $M_{2}$ and $M_{3}$ can be \emph{simultaneously}
(jointly) performed. Denote the joint measurement by $M_{1}\wedge M_{2}\wedge M_{3}$
and the eight possible joint outcomes by $O_{1}^{\pm}\wedge O_{2}^{\pm}\wedge O_{3}^{\pm}$.
For this simultaneous measurement quantum mechanics predicts the following
distribution of the joint outcomes:

\begin{eqnarray}
\braket{\Psi\vert P_{1}^{\pm}P_{2}^{\pm}P_{3}^{\pm}\Psi} & = & p_{\Psi}(O_{1}^{\pm}\wedge O_{2}^{\pm}\wedge O_{3}^{\pm}\,\vert\,M_{1}\wedge M_{2}\wedge M_{3})\label{eq:QM2}
\end{eqnarray}

It turns out that only four of the eight joint outcomes have nonzero
probability: those that correspond to the four common eigenvectors
(Bell state vectors) of the three operators:
\begin{center}
\begin{tabular}{c|ccc}
{\footnotesize{}{}}%
\mbox{%
{\footnotesize{}{}Eigenvectors and eigenstates}%
}{\footnotesize{}{}} & {\footnotesize{}{}$A_{1}$} & {\footnotesize{}{}$A_{2}$} & {\footnotesize{}{}$A_{3}$}\tabularnewline
\hline 
{\footnotesize{}{}$\ket{\Psi^{+-+}}=\frac{1}{\sqrt{2}}\big(\ket{00}+\ket{11}\big)$} & {\footnotesize{}{}$+1$} & {\footnotesize{}{}$-1$} & {\footnotesize{}{}$+1$}\tabularnewline
{\footnotesize{}{}$\ket{\Psi^{-++}}=\frac{1}{\sqrt{2}}\big(\ket{00}-\ket{11}\big)$} & {\footnotesize{}{}$-1$} & {\footnotesize{}{}$+1$} & {\footnotesize{}{}$+1$}\tabularnewline
{\footnotesize{}{}$\ket{\Psi^{++-}}=\frac{1}{\sqrt{2}}\big(\ket{01}+\ket{10}\big)$} & {\footnotesize{}{}$+1$} & {\footnotesize{}{}$+1$} & {\footnotesize{}{}$-1$}\tabularnewline
{\footnotesize{}{}$\ket{\Psi^{---}}=\frac{1}{\sqrt{2}}\big(\ket{01}-\ket{10}\big)$} & {\footnotesize{}{}$-1$} & {\footnotesize{}{}$-1$} & {\footnotesize{}{}$-1$}\tabularnewline
\end{tabular}
\par\end{center}

\noindent where $\ket{0}$ and $\ket{1}$ are the eigenvectors of
$\sigma_{z}$ with eigenvalue $+1$ and $-1$. This means that the
simultaneous measurement $M_{1}\wedge M_{2}\wedge M_{3}$ can have
only four possible joint outcome types in every quantum state:
\begin{equation}
O_{1}^{+}\wedge O_{2}^{-}\wedge O_{3}^{+}\quad,\quad\quad O_{1}^{-}\wedge O_{2}^{+}\wedge O_{3}^{+}\quad,\quad\quad O_{1}^{+}\wedge O_{2}^{+}\wedge O_{3}^{-}\quad,\quad\quad O_{1}^{-}\wedge O_{2}^{-}\wedge O_{3}^{-}\label{eq:OOO}
\end{equation}

\noindent Consequently, only four possible joint outcome types are
allowed in every hidden state.

Now, let $(\Lambda,p)$ be a noncontextual value-definite hidden-variable
model for a realization of $A_{1},$ $A_{2}$ and $A_{3}$ by the
simultaneous measurements of $M_{1}$, $M_{2}$ and $M_{3}$. The
set $\Lambda$ is composed of four types of hidden states: 
\begin{equation}
\Lambda=\Lambda^{+-+}\cup\Lambda^{-++}\cup\Lambda^{++-}\cup\Lambda^{---}\label{eq:LLL}
\end{equation}
 corresponding to the above four joint outcomes. For example, for
a system in a hidden state in $\Lambda^{+-+}$, the outcome of the
simultaneous measurement $M_{1}\wedge M_{2}\wedge M_{3}$ will be
$O_{1}^{+}\wedge O_{2}^{-}\wedge O_{3}^{+}$ and the outcome of the
individual measurements $M_{1}$, $M_{2}$ and $M_{3}$ will be $O_{1}^{+}$,
$O_{2}^{-}$ and $O_{3}^{+}$, respectively. Thus, the model is value-definite
and noncontextual. The probability of the four hidden state types
for a system in quantum state $\ket{\Psi}$ is

\begin{eqnarray}
p\left(\Lambda^{+-+}\right) & = & \vert\braket{\Psi\vert\Psi^{+-+}}\vert^{2}\label{eq:HV1a}\\
p\left(\Lambda^{-++}\right) & = & \vert\braket{\Psi\vert\Psi^{-++}}\vert^{2}\\
p\left(\Lambda^{++-}\right) & = & \vert\braket{\Psi\vert\Psi^{++-}}\vert^{2}\\
p\left(\Lambda^{---}\right) & = & \vert\braket{\Psi\vert\Psi^{---}}\vert^{2}\label{eq:HV1d}
\end{eqnarray}

\noindent Since the eigenvectors form an orthonormal basis, the probabilities
add up to 1. Furthermore, from

\noindent 
\begin{eqnarray}
\vert\braket{\Psi\vert\Psi^{+-+}}\vert^{2} & = & \braket{\Psi\vert P_{1}^{+}P_{2}^{-}P_{3}^{+}\Psi}\label{eq:eigen1}\\
\vert\braket{\Psi\vert\Psi^{-++}}\vert^{2} & = & \braket{\Psi\vert P_{1}^{-}P_{2}^{+}P_{3}^{+}\Psi}\\
\vert\braket{\Psi\vert\Psi^{++-}}\vert^{2} & = & \braket{\Psi\vert P_{1}^{+}P_{2}^{+}P_{3}^{-}\Psi}\\
\vert\braket{\Psi\vert\Psi^{---}}\vert^{2} & = & \braket{\Psi\vert P_{1}^{-}P_{2}^{-}P_{3}^{-}\Psi}\label{eq:eigen4}
\end{eqnarray}
and (\ref{eq:QM2}) it follows that 

\begin{eqnarray}
p\left(\Lambda^{+-+}\right) & = & p_{\Psi}(O_{1}^{+}\wedge O_{2}^{-}\wedge O_{3}^{+}\,\vert\,M_{1}\wedge M_{2}\wedge M_{3})\label{eq:HV1a-1}\\
p\left(\Lambda^{-++}\right) & = & p_{\Psi}(O_{1}^{-}\wedge O_{2}^{+}\wedge O_{3}^{+}\,\vert\,M_{1}\wedge M_{2}\wedge M_{3})\\
p\left(\Lambda^{++-}\right) & = & p_{\Psi}(O_{1}^{+}\wedge O_{2}^{+}\wedge O_{3}^{-}\,\vert\,M_{1}\wedge M_{2}\wedge M_{3})\\
p\left(\Lambda^{---}\right) & = & p_{\Psi}(O_{1}^{-}\wedge O_{2}^{-}\wedge O_{3}^{-}\,\vert\,M_{1}\wedge M_{2}\wedge M_{3})\label{eq:HV1d-1}
\end{eqnarray}

Next, suppose that the three measurements realizing the operators
cannot be performed simultaneously but only \emph{sequentially}. This
means that the constraint (\ref{eq:QM2}) does not apply to the measurements
for the simple reason that the right hand side is not defined. There
is, however, another constraint coming from the projection postulate.
According to the projection postulate, upon performing the measurement
$M_{i}$ on the system and getting the outcome $O_{i}^{\pm}$, the
quantum state $\ket{\Psi}$ will jump into the new state 
\begin{equation}
\ket{\Psi}\quad\stackrel{}{\longrightarrow}\quad\ket{\Psi'}=\frac{P_{i}^{\pm}\ket{\Psi}}{\braket{\Psi\vert P_{i}^{\pm}\Psi}^{\frac{1}{2}}}\label{eq:PP}
\end{equation}

Now, suppose we perform the three measurement $M_{1}$, $M_{2}$ and
$M_{3}$ one after another and obtain an outcome for each measurement.
The first measurement will send the system into a new quantum state,
$\ket{\Psi'}$; the second measurement will send it further into another
quantum state, $\ket{\Psi''}$ according to the projection postulate.
We now ask: What is the probability that we obtain a given sequence
of outcomes upon these sequential measurements? A simple calculation
shows that this probability is just the quantum probability on the
left hand side of (\ref{eq:QM2}):

\begin{equation}
p_{\Psi}(O_{1}^{\pm}\,\vert\,M_{1})\cdot p_{\Psi'}(O_{2}^{\pm}\,\vert\,M_{2})\cdot p_{\Psi''}(O_{3}^{\pm}\,\vert\,M_{3})=\braket{\Psi\vert P_{1}^{\pm}P_{2}^{\pm}P_{3}^{\pm}\Psi}\label{eq:QM3}
\end{equation}

\noindent It is easy to show that the right hand side of (\ref{eq:QM3})
remains the same even if the measurements are performed in a different
order. Therefore, we introduce a general notation for an arbitrary
sequence of measurements $M_{1}$, $M_{2}$ and $M_{3}$ performed
on a system in quantum state $\ket{\Psi}$ with outcomes $O_{1}^{\pm}$,
$O_{2}^{\pm}$ and $O_{3}^{\pm}$:

\begin{equation}
p_{\Psi}(O_{1}^{\pm}\!-\!O_{2}^{\pm}\!-\!O_{3}^{\pm}\,\vert\,M_{1}\!-\!M_{2}\!-\!M_{3})=\braket{\Psi\vert P_{1}^{\pm}P_{2}^{\pm}P_{3}^{\pm}\Psi}\label{eq:QM4}
\end{equation}

\noindent Note that while $M_{1}\wedge M_{2}\wedge M_{3}$ denoted
the simultaneous measurement of $M_{1}$, $M_{2}$ and $M_{3}$, the
term $M_{1}\!-\!M_{2}\!-\!M_{3}$ denotes a sequential measurement
of $M_{1}$, $M_{2}$ and $M_{3}$ in any order.

(\ref{eq:QM4}) means that the sequential measurement $M_{1}\!-\!M_{2}\!-\!M_{3}$
can yield only those sequential outcomes which conform to the eigenvalues
of one of the above four common eigenvectors. Moreover, the probabilities
of these sequential outcomes will be just the probabilities in (\ref{eq:eigen1})-(\ref{eq:eigen4}).
Thus, (\ref{eq:QM4}) together with the projection postulate provide
another interpretation of the term $\braket{\Psi\vert P_{1}^{\pm}P_{2}^{\pm}P_{3}^{\pm}\Psi}$:
it will no longer represent the joint outcomes of the simultaneous
measurement $M_{1}\wedge M_{2}\wedge M_{3}$, as in (\ref{eq:QM2});
rather it will represent the outcomes of the sequential measurement
$M_{1}\!-\!M_{2}\!-\!M_{3}$, as in (\ref{eq:QM4}). 

Now, what kind of noncontextual value-definite hidden-variable models
are admitted if the operators $A_{1}$, $A_{2}$ and $A_{3}$ are
realized not by simultaneous but by sequential measurements conforming
to (\ref{eq:QM4})?

Obviously, the previous noncontextual value-definite hidden-variable
model is a hidden-variable model also for this sequential realization
of the operators. Here measurements do not change the hidden state
of the system. By performing a measurement and selecting out those
runs which yield a given outcome, one changes only the distribution
of the hidden states. This change will be consistent with the change
of the quantum state via the projection postulate (\ref{eq:PP}),
that is the new distribution can be calculated by replacing $\ket{\Psi}$
with $\ket{\Psi'}$ in (\ref{eq:HV1a})-(\ref{eq:HV1d}). 

There is, however, another noncontextual value-definite hidden-variable
model for this sequential realization of the operators where the joint
values do \emph{not} correspond to the above four eigenvectors. Let
now $\Lambda$ be composed of eight types of hidden states $\Lambda^{\pm\pm\pm}$:
four corresponding to the above four joint outcomes and four corresponding
to the joint outcomes with opposite signs. Let the probability of
hidden state types for a system in quantum state $\ket{\Psi}$ be

\begin{eqnarray}
p\left(\Lambda^{\pm\pm\pm}\right) & = & \braket{\Psi\vert P_{1}^{\pm}\Psi}\cdot\braket{\Psi\vert P_{2}^{\pm}\Psi}\cdot\braket{\Psi\vert P_{3}^{\pm}\Psi}\label{eq:HV2}
\end{eqnarray}

Note that in this model the joint outcome of the simultaneous measurement
$M_{1}\wedge M_{2}\wedge M_{3}$ in the hidden state $\Lambda^{\pm\pm\pm}$
of the system \emph{would} be $O_{1}^{\pm}\wedge O_{2}^{\pm}\wedge O_{3}^{\pm}$.
However, $M_{1}$, $M_{2}$ and $M_{3}$ cannot be simultaneously
measured, only sequentially and individually. If measured individually,
the measurements provide the probabilities consistent with quantum
mechanics:

\[
\sum_{jk=\pm}p\left(\Lambda^{\pm jk}\right)=\braket{\Psi\vert P_{1}^{\pm}\Psi}=p_{\Psi}(O_{1}^{\pm}\,\vert\,M_{1})
\]

\noindent and similarly for $i=2,3$. The model is noncontextual since
the observables associated with $M_{1}$, $M_{2}$ and $M_{3}$ have
joint values, even if these measurements cannot be performed simultaneously.

Also note that the absence of simultaneous measurements is crucial:
the hidden state type $\Lambda^{+++}$, for example, which was ruled
out in the previous model since the joint outcome $O_{1}^{+}\wedge O_{2}^{+}\wedge O_{3}^{+}$
could never pop up for a simultaneous measurement, is \emph{not} ruled
out here. If the system is prepared in the quantum state $\ket{\Psi}=\ket{00}$,
for example, then the probability of the type $\Lambda^{+++}$ is
nonzero:
\[
p\left(\Lambda^{+++}\right)=\braket{00\vert P_{1}^{+}00}\cdot\braket{00\vert P_{2}^{+}00}\cdot\braket{00\vert P_{3}^{+}00}=\frac{1}{4}
\]

Our task is now to make the model consistent also diachronically with
respect to the sequential measurements. In other words, we need to
introduce a change in the distribution of the hidden states which
is consistent with (\ref{eq:QM4}). One can reach this goal by ensuring
that the distribution of hidden states upon every measurement conforms
to the updated quantum state. If the change of the distribution $p\rightarrow p'\rightarrow p''\rightarrow...$
of the hidden states upon a sequence of measurements is consistent
with the change of the quantum states $\ket{\Psi}\rightarrow\ket{\Psi'}\rightarrow\ket{\Psi''}\rightarrow...$
as governed by the projection postulate---that is probabilities relate
to the quantum states via (\ref{eq:HV2}) in every step of the measurement
process, then (\ref{eq:QM3}) and hence (\ref{eq:QM4}) will hold
trivially. 

There are many stochastic transition processes which satisfy this
requirement (see La Cour, 2009, 2017; Kleinmann et al., 2011; Cabello
et al., 2018). Here is the presumably simplest (and admittedly a least
realistic) one: upon performing the measurement $M_{i}$ on the system
and getting the outcome $O_{i}^{\pm}$, let each hidden state in $\Lambda^{\pm\pm\pm}$
jump into a new hidden state in $\Lambda'^{\pm\pm\pm}$: 
\begin{equation}
\lambda\in\Lambda^{\pm\pm\pm}\quad\stackrel{p'}{\longrightarrow}\quad\lambda'\in\Lambda'^{\pm\pm\pm}\label{eq:PP2}
\end{equation}

\noindent with probability $p'\left(\Lambda'^{\pm\pm\pm}\right)$---that
is with the probability of the new hidden state type $\Lambda'^{\pm\pm\pm}$
calculated by (\ref{eq:HV2}) with $\ket{\Psi}$ replaced by $\ket{\Psi'}$. 

As an example, consider a system in a hidden state in $\Lambda{}^{+++}$
which is measured sequentially by $M_{1}$, $M_{2}$ and $M_{3}$.
The outcome of $M_{1}$ will be $O_{1}^{+}$ and the hidden state
will jump into one of the four types $\Lambda{}^{+jk}$ with probability
$\braket{\Psi'\vert P_{2}^{j}\Psi'}\cdot\braket{\Psi'\vert P_{3}^{k}\Psi'}$,
where

\[
\ket{\Psi'}=\frac{P_{1}^{+}\ket{\Psi}}{\braket{\Psi\vert P_{1}^{+}\Psi}^{\frac{1}{2}}}
\]
Suppose the system remains in $\Lambda{}^{+++}$ after the first measurement.
If we perform the second measurement $M_{2}$ on this system, the
outcome will be $O_{2}^{+}$ and the system will jump into $\Lambda{}^{++-}$
with probability 1 (since the new projected state $\ket{\Psi''}$
is just $\ket{\Psi^{++-}}$.) In this new hidden state, the third
measurement $M_{3}$ will give the outcome $O_{3}^{-}$. 

This example highlights a general rule which the jumps need to follow.
Independent of which hidden state the system starts from, after two
consecutive measurements it will land in one of the four states in
(\ref{eq:LLL}) corresponding to the joint outcomes. This must be
so since repeating any of the two measurements, the outcome needs
to be same as before; and performing the third measurement, the outcome
needs to be one of (\ref{eq:OOO}). Thus, hidden states not in (\ref{eq:LLL})
are washed out after two sequential measurements. We come back to
this point in the Discussion.

Also note that the probability $p'$ of the transition between the
old and new hidden states in (\ref{eq:PP2}) depends on the new quantum
state $\ket{\Psi'}$ via (\ref{eq:HV2}). But this new quantum state
$\ket{\Psi'}$ is determined by the old quantum state $\ket{\Psi}$,
the measurement $M_{i}$ and the outcome $O_{i}^{\pm}$ via (\ref{eq:PP}).
Therefore, to correctly govern the transition, the hidden-variable
model needs to incorporate also the quantum states $\ket{\Psi}$.
Thus, the hidden states will be of the form $\left\{ \lambda,\ket{\Psi}\right\} $,
where $\lambda\in\Lambda$ and $\ket{\Psi}\in H_{2}$. In short, the
model will be \emph{$\Psi$-ontic} (see Harrigan and Spekkens, 2010).

In the next section, I construct a similar noncontextual value-definite
hidden-variable model with stochastic transitions for the sequential
realization of the operators in the Peres-Mermin square.

\section{The Peres-Mermin square }

The Peres-Mermin square (Peres, 1990; Mermin 1993) is the following
$3\!\times\!3$ matrix of self-adjoint operators: 
\begin{center}
\begin{tabular}{ccc}
\quad{}$A_{11}=\sigma_{z}\otimes I$ \quad{} & \quad{}$A_{12}=I\otimes\sigma_{z}$ \quad{} & \quad{}$A_{13}=\sigma_{z}\otimes\sigma_{z}$ \quad{}\tabularnewline
 &  & \tabularnewline
\quad{}$A_{21}=I\otimes\sigma_{x}$ \quad{} & \quad{}$A_{22}=\sigma_{x}\otimes I$ \quad{} & \quad{}$A_{23}=\sigma_{x}\otimes\sigma_{x}$ \quad{}\tabularnewline
 &  & \tabularnewline
\quad{}$A_{31}=\sigma_{z}\otimes\sigma_{x}$ \quad{} & \quad{}$A_{32}=\sigma_{x}\otimes\sigma_{z}$ \quad{} & \quad{}$A_{33}=\sigma_{y}\otimes\sigma_{y}$ \quad{}\tabularnewline
\end{tabular}
\par\end{center}

\noindent where $I$ is the unit operator on $H_{2}$. Each operator
in the matrix has two eigenvalues, $\pm1$, and are arranged in such
a way that two operators are commuting if and only if they are in
the same row or in the same column. The three operators in the third
column are just the three commuting operators in the previous section.

A realization (interpretation) of the Peres-Mermin square is a unique
association of operators $\left\{ A_{ij}\right\} $ $i,j=1,2,3$ in
the matrix with real-world measurements $\left\{ M_{ij}\right\} $.
Suppose that the measurements realizing commuting operators cannot
be simultaneously performed but only sequentially. In other words,
instead of performing the joint measurement $M_{1j}\wedge M_{2j}\wedge M_{3j}$
one can only perform the sequential measurements $M_{1j}\!-\!M_{2j}\!-\!M_{3j}$
. Similarly, instead of performing the joint measurement $M_{i1}\wedge M_{i2}\wedge M_{i3}$
one can only perform the sequential measurements $M_{i1}\!-\!M_{i2}\!-\!M_{i3}$.
The realization is empirically adequate if (\ref{eq:QM1}) and (\ref{eq:QM4})
hold for the individual and sequential measurements.

I will construct now a noncontextual value-definite hidden-variable
model for this realization of the Peres-Mermin square. The model will
recover the outcome statistics of both the individual measurements
and the sequential measurements. Still, the joint values of the observables
corresponding to the commuting triples of operators will \emph{not}
correspond to the eigenvalues in one of the common eigenstates of
these operators. Consequently, the usual constraints on the valuations
leading to the Kochen-Specker contradiction will not apply. The model
is not ruled out by the Kochen-Specker arguments.

Let $\varepsilon=(\varepsilon_{11},\varepsilon_{12},\dots,\varepsilon_{33})$
be a vector such that $\varepsilon_{ij}=\pm1$. Let $P_{ij}^{\varepsilon_{ij}}$
denote the eigenprojection of the operator $A_{ij}$ with eigenvalue
$\varepsilon_{ij}$ and let $\ket{\Psi}$ be the quantum state of
the system. The hidden-variable model consist of hidden states $\left\{ \lambda,\ket{\Psi}\right\} \in\Lambda\times H_{2}$
where $\Lambda$ is composed of $2^{9}$ hidden state types:
\begin{eqnarray*}
\Lambda & = & \bigcup_{\varepsilon\in\{-1,+1\}^{9}}\Lambda^{\varepsilon}
\end{eqnarray*}

\noindent such that $\lambda\in\Lambda^{\varepsilon}$ if and only
if the outcome of the measurement $M_{ij}$ would be $O_{ij}^{\varepsilon_{ij}}$
for $i,j=1,2,3$. The probability of the hidden state type $\Lambda^{\varepsilon}$
is 
\begin{eqnarray}
p(\Lambda^{\varepsilon}) & = & \prod_{ij}p_{ij}^{\varepsilon_{ij}}\quad\quad\quad\mbox{where}\quad p_{ij}^{\varepsilon_{ij}}=\braket{\Psi\vert P_{ij}^{\varepsilon_{ij}}\Psi}\label{eq:HV3}
\end{eqnarray}
The probability of those hidden states which provide the outcome $O_{ij}^{\pm}$
for the measurement $M_{ij}$ is 
\begin{equation}
\sum_{\varepsilon:\,\varepsilon_{ij}=\pm1}p(\Lambda^{\varepsilon})=p_{ij}^{\pm1}\label{eq:HV4}
\end{equation}
The probabilities are normalized: 
\[
\sum_{\varepsilon}p(\Lambda^{\varepsilon})=1
\]

Now, according to the projection postulate (\ref{eq:PP}), upon performing
the measurement $M_{ij}$ on the system and getting the outcome $O_{ij}^{\pm}$,
the quantum state $\ket{\Psi}$ jumps into the new state 
\[
\ket{\Psi}\quad\stackrel{}{\longrightarrow}\quad\ket{\Psi'}=\frac{P_{ij}^{\pm}\ket{\Psi}}{\braket{\Psi\vert P_{ij}^{\pm}\Psi}{}^{\frac{1}{2}}}
\]
For the model to be consistent, the new probability distribution of
the hidden states after the measurement should be given again by (\ref{eq:HV3})
with $\ket{\Psi}$ replaced by $\ket{\Psi'}$. This can be guaranteed
again by the following simple stochastic transition process between
the old and new hidden states: upon performing the measurement $M_{ij}$
and getting the outcome $O_{ij}^{\pm}$, each hidden state in $\Lambda^{\varepsilon}$
jumps into the new hidden state type $\Lambda'^{\varepsilon}$ with
probability $p(\Lambda'^{\varepsilon})$. This transition guarantees
that the distribution of hidden states will co-vary with the quantum
state in tune with the projection postulate. Specifically, the probability
of those new hidden states which provide again the outcome $O_{ij}^{\pm}$
if we repeat the measurement $M_{ij}$ will be 
\[
\sum_{\varepsilon:\,\varepsilon_{ij}=\pm1}p(\Lambda'^{\varepsilon})=1
\]
since $\bra{\Psi'}P_{ij}^{\pm1}\ket{\Psi'}=1$.

Note that the probability $p(\Lambda'^{\varepsilon})$ of the transition 

\begin{equation}
\lambda\in\Lambda^{\varepsilon}\quad\stackrel{p'}{\longrightarrow}\quad\lambda'\in\Lambda'^{\varepsilon}\label{eq:PP3}
\end{equation}
does not depend on the old probability $p(\Lambda^{\varepsilon})$
but it does depend on $\ket{\Psi'}$ which is determined by the old
quantum state $\ket{\Psi}$, the measurement $M_{ij}$ and the outcome
$O_{ij}^{\pm}$. This is why the hidden-variable model needs to incorporate
also the quantum state $\ket{\Psi}$ which makes the model $\Psi$-ontic.
There is a division of labor in the model: the $\lambda$-part of
the hidden states $\left\{ \lambda,\ket{\Psi}\right\} $ ensures that
the individual measurements have definite outcomes; the $\ket{\Psi}$-part
governs (stochastically) the update of the hidden states upon measurements.
Note that since no measurements can be simultaneously performed, the
model does not make a difference between measurements realized by
commuting and noncommuting operators: it treats them alike.

To sum up, if the Peres-Mermin square is not realized by simultaneous
but only by sequential measurements, then it does not rule out a value-definite
(and $\Psi$-ontic) hidden variable model respecting also the projection
postulate. 

\section{Two recent experiments with sequential measurements}

In a recent experiment, Kirchmair et al. (2009) realized the Peres-Mermin
operators by sequential measurements performed on pairs of $^{40}$Ca$^{+}$
ions. In the experiment, trapped ions were prepared in a two-qubit
quantum state by laser--ion interactions. The observable associated
with the operator $\sigma_{z}$ was realized by two different energy
levels of the ions. This observable was measured by electron shelving
projecting onto these eigenstates. The measurement of the other eight
observables in the Peres-Mermin square was reduced to the measurement
of the observable represented by $\sigma_{z}$ by applying a suitable
unitary transformation to the quantum state before this measurement
and its inverse after this measurement. Thus, the association of the
operators and measurements was unique: each of the 9 operators was
associated with a different (quantum non-demolition) measurement. 

The aim of the experiment was to test the violation of the Peres-Mermin
inequality 
\begin{equation}
\langle A_{11}A_{12}A_{13}\rangle+\langle A_{21}A_{22}A_{23}\rangle+\langle A_{31}A_{32}A_{33}\rangle+\langle A_{11}A_{21}A_{31}\rangle+\langle A_{21}A_{22}A_{32}\rangle-\langle A_{13}A_{23}A_{33}\rangle\leqslant4\label{eq:PMineq}
\end{equation}

\noindent derived by Cabello (2008) as a constraint on the Peres-Mermin
square to have a noncontextual value-definite hidden variable model.
For the right hand side of (\ref{eq:PMineq}), quantum mechanics predicts
6 in any quantum state which violates the inequality. The experiment
of Kirchmair et al. confirmed this prediction by obtaining the result
5.46 for the singlet state.

The violation of (\ref{eq:PMineq}), however, does not rule out a
noncontextual value-definite hidden variable model the Peres-Mermin
square since the measurements realizing the three operators in a row
or column are not simultaneously performed. Just consider the model
developed in the previous section. For every quantum state, (\ref{eq:HV3})
provides the probability distribution of the hidden state types which
returns the outcome statistics of the individual measurements via
(\ref{eq:HV4}). Upon performing a measurement and obtaining an outcome,
the quantum state updates in tune with the projection postulate and
the hidden states stochastically jump into another type with the probability
of this new type. The expectation value of the subsequent measurement
of three observables in a row or column will provide just the six
expressions on the right hand side of (\ref{eq:PMineq}) leading to
the violation of the inequality. Still, the model is noncontextual
in every step of the measurement process: the nine observables have
joint values at each moment.

The experiment of Kirchmair et al. has been further developed and
carried out by photons by Liu et al. (2016). In this experiment, two
entangled photons were distributed between two spatially separated
parties, Alice and Bob. Both photons encoded two qubits, one in the
spatial and another in the polarization mode. Thus, the photon pair
was in a four-qubit quantum state. Alice performed three sequential
measurements on her photon and Bob performed one single measurement
on his photon. The sequential measurements of Alice realized one of
the three rows or columns of the Peres-Mermin square using beam splitters,
half-wave plates, beam displacers and phase compensators. Bob's single
measurement realized one of the Peres-Mermin operator. Now, in tune
with quantum mechanics, if Bob chose a measurement which was identical
with the second or third measurement in Alice's sequence of measurements,
then there was a perfect correlation or anticorrelation between their
outcomes. 

The aim of the experiment was to verify the violation of a generalization
by Cabello's (2010) of the Peres-Mermin inequality (\ref{eq:PMineq})
where the perfect correlation or anticorrelation terms were added.
The experiment proved the violation of this generalized Peres-Mermin
inequality. 

However, similarly to the experiment of Kirchmair et al., the experiment
of Liu et al. does not rule out noncontextual value-definite hidden
variable models. A sophisticated model for the experiment of Liu et
al. was given by La Cour (2017). But more simple-minded models can
also be given. Here we just sketch how it goes: Quantum mechanics
determines the quantum state of the system after each measurement
via the projection postulate. Use (\ref{eq:HV3}) to establish the
probability of the hidden states of Alice in every step of the measurement
process and use the perfect correlation and anticorrelation between
the outcomes of Alice's and Bob's measurement to establish the probability
of the hidden states of Bob.

Instead of continuing with other more recent experiments (see i.e.
Leupold et al., 2018) let me make once more explicit the crucial difference
between simultaneous and sequential measurements. If a set of measurements
can be performed only sequentially, then only one measurement can
be performed at a time on the system. In this case the hidden variable
model needs to return only the outcome statistics of the \emph{individual}
measurements. If, however, the measurements can be performed also
simultaneously, then the hidden variable model needs to provide the
statistics of both the \emph{individual} and the \emph{simultaneous}
measurements; moreover, to be noncontextual, it needs to yield the
same outcomes. 

\section{Discussion}

With this paper, I intend to contribute to the debate on the experimental
testability of the Kochen-Specker theorems via \emph{sequential} measurements.
One central question of this debate is whether the standard notion
of noncontextuality is applicable in case of sequential measurements
or one needs to adapt the concept to these new experimental conditions. 

Some authors opt for the second alternative. Gühne et al. (2010, Def.
2), for example, calls a hidden variable model noncontextual if in
a hidden state the outcome of a measurement does not depend on whether
another compatible measurement---represented by a commuting operator---is
measured \emph{before} it, \emph{simultaneously} with it, or \emph{after}
it. In other words, if we perform a measurement on a system in a given
hidden state and obtain an outcome, this outcome would be the same
had we performed a compatible measurement or even a whole sequence
of compatible measurements before or jointly with or after it. For
compatible measurements, the outcomes are fixed once and for all and
are not sensitive to the order of measurements. 

In their paper, Gühne et al. analyze the additional assumptions leading
to the violation of Kochen-Specker inequalities in case of sequential
measurements. One such assumption is what they call "compatibility
loophole''. The authors investigate the possibility of abandoning
\emph{perfect compatibility,} that is, to allow for a measurement
to provide, at least sometimes, different outcomes in a sequence of
compatible measurements. They show how this \emph{imperfect compatibility}
can lead to different modifications of the Kochen-Specker inequalities;
compare these inequalities with real-world experiments; and construct
various contextual hidden variable models.

The present paper differs from that of Gühne et al. in two important
points. First, the model developed in Section 3 satisfies perfect
compatibility. The stochastic transition (\ref{eq:PP3}) was explicitly
designed such that it tracks the transformation of the wave function
under the projection postulate. Upon any sequence of measurements
corresponding to a given row or column of the Peres-Mermin square,
the outcomes will always conform to the one of the four common eigenstates
of the three commuting operators in that row or column. I also showed
that since after two consecutive measurements the quantum state will
be projected onto one of the common eigenstates, the hidden state
of the system will be the state corresponding to these joint outcomes.
Only after performing a measurement realizing a \emph{non}-commuting
operator can the system leave this hidden state.

Second, I opted for the first alternative in the above dilemma and
sticked with the traditional definition of noncontextuality. I called
a hidden variable model noncontextual if the observables associated
with the commuting operators have joint values in every hidden state.
These joint values can be revealed only by \emph{simultaneous} measurements
and the model is noncontextual only if these values do not depend
on whether the simultaneous measurements are performed or not. Noncontextuality,
in my understanding, does not include that measurements cannot alter
the hidden state of the system and hence cannot alter the outcome
of a subsequent measurement represented by a commuting operator. In
a common eigenstate of two operators this is certainly the case: both
measurements have a fixed value and these values remain the same no
matter how many times and in what order we perform the measurements.
But generally, in a non-eigenstate noncontextuality as defined by
Gühne et al. seems to be too strong: observables can well have joint
values at each time which values update for every new measurement. 

La Cour (2009) also rejects the definition of noncontextuality as
defined by Gühne et al. As he writes: "In the broadest sense, a
measurement of an observable is said to be noncontextual if the outcome
of the measurement does not depend upon which other compatible observables
are measured subsequently, simultaneously, or previously... A better
definition of a noncontextual measurement, then, would require only
that the joint statistics of commuting observables be unchanged by
the details of how they are measured.'' (p. 012102-1) And he goes
on and constructs a value-definite, noncontextual hidden-variable
model which reproduces the quantum statistics of the Mermin-Peres
square. The model is highly sophisticated. It specifies the change
of the hidden states upon sequential and simultaneous measurements;
satisfies perfect compatibility; and provides a model for certain
recent real-world photon and neutron interferometry experiments. The
restrictions of the standard Kochen-Specker theorems are avoided in
the same way as in this paper: by allowing for the hidden states to
change during the measurements. 

La Cour's definition of noncontextuality, however, is different from
the concept of noncontextuality used in this paper. Noncontextuality
in La Cour is a statistical feature of the model in the sense that
one associates a single random variable with each operator in the
Mermin-Peres square and reproduces the quantum statistics with different
probability measures corresponding to the different experiments. Although
La Cour's model provides a deterministic mechanism for the update
of hidden states upon measurements, his model qualities as \emph{contextual}
in our terminology. The reason for that lies in the difference how
the two models treat the \emph{order} of measurement outputs and measurement
interactions. In our model a given hidden state \emph{first} determines
the measurement outcome and \emph{then} updates due to the measurement
interaction. In La Cour's model the order is just the opposite: an
initial hidden state \emph{first} transforms into a new state depending
on the chosen measurement and \emph{then} this new hidden state determines
the outcomes. Obviously, one can argue for either order---being both
experimentally inaccessible (see Conclusions). Interestingly, however,
La Cour's choice of order makes his model contextual, at least with
respect to our terminology. Namely, in La Cour two compatible measurements,
say, $M_{11}$ and $M_{11}\wedge M_{12}\wedge M_{13}$ will take an
initial hidden state $\lambda$ into two different new states $\lambda'$
and $\lambda''$ which \emph{then} can yield different outcomes. La
Cour does not consider this feature contextually. He writes: "As
discussed previously, this is not a violation of noncontextuality
but merely a reflection of the possible dependence of a particular
outcome on the experimental procedure.'' (p. 012102-1) In our model,
however, $\lambda$ first determines the outcome for both measurements
and then updates according the projection postulate. Since the outcome
for $M_{11}$ and the would-be outcome\footnote{Note that the joint measurement $M_{11}\wedge M_{12}\wedge M_{13}$
cannot be performed; but if it could, it would yield a definite outcome
in every hidden state.} for $M_{11}\wedge M_{12}\wedge M_{13}$ (with respect to $M_{11}$),
we call the model noncontextual. This shows that the order of output
and update during measurements is strongly connected to the concept
of noncontextuality. To make La Cour's model noncontextual with respect
to our terminology, one should also demand that compatible measurements
drive the hidden states into the same new state (or at least into
such different states for which the outcome of the common part of
the measurements---in this case $M_{11}$---is the same).

Another difference is that at certain points of his paper, La cour
seems to make concessions to the traditional wording by referring
to the dependence of the random variables or the probability measures
on the commuting sets as "effective'' or "apparent contextuality''
and explaining this contextuality by the interaction of the system
with the measurement apparatus. We, however, reserve the term "contextual''
only to the dependence of a measurement outcome in a given hidden
state on whether another measurement is performed on the same system
at the same time. As stressed above, I think that it is worth discerning
1) the\emph{ robustness of the system to respond} in a definite way
to a measurement when simultaneous measurements are also performed
and 2) the \emph{robustness of the hidden state to change} when it
interacts with a measurement apparatus.

\section{Conclusions}

In the paper I argued that the Peres-Mermin square does not rule out
a value-definite noncontextual hidden variable model if the observables
associated with the operators cannot be measured simultaneously but
only sequentially. To highlight this claim, I constructed such a model
for any realization of the Peres-Mermin square by sequential measurements. 

I would like to conclude with two remarks. 
\begin{enumerate}
\item The stochastic transition of the hidden states upon measurement is
\emph{not necessarily local} in the sense that if $M_{i}$ and $M_{j}$
are measurements on two spacelike separated subsystems, then upon
measuring $M_{i}$ and obtaining an outcome, the hidden state $\lambda{}^{\varepsilon}$
can jump into a hidden state $\lambda'{}^{\varepsilon}$ for which
the outcome of measurement $M_{j}$ will be different from that in
$\lambda{}^{\varepsilon}$. But this nonlocal character of the transition
is a different feature of the theory than noncontextuality as the
robustness of the outcome of a measurement against simultaneous measurements.
The aim of this paper is not to dispute the claim that quantum mechanics
does not admit local hidden variable models---we know this from Bell's
inequalities. The aim is simply to argue that the Kochen-Specker arguments
do not rule out a value-definite noncontextual hidden variable model
if the observables associated with the operators can be measured only
sequentially. 
\item When are two measurements simultaneous and when are they subsequent?
What time difference is needed for two measurements to be sequential?
Well, this question cannot be answered a priori. It depends on the
nature of the interaction between the system and the measuring apparatus.
Still, there is a conceptual difference between simultaneous and subsequent
measurements. In case of sequential measurements, the first measurement
has time, so to say, to alter the hidden state and hence to influence
the system's response to a second measurement. In the experiments
analyzed above this is clearly the case: photons when entering a measurement
apparatus and when leaving it need not be in the same hidden state
due to their interaction with the apparatus. Consequently, they can
enter the subsequent measurement apparatus in a new hidden state.
As said above, I do not call this phenomenon contextuality. Contextuality
occurs only when a measurement has a direct causal influence on the
outcome of another measurement without previously altering the hidden
state of the system. And our best way to ensure this, is to perform
the two measurements simultaneously. 

The sceptic might respond: Simultaneous and sequential measurements
are on a par since neither simultaneous measurements can completely
rule out that the hidden states update between the measurements. Since
measurements take time, it can well happen that the system's interaction
with the one apparatus happens much faster than with the other and
hence the hidden state can update between the two measurements. In
this case, an updating noncontextual hidden variable model could be
provided also for such simultaneous measurements. This is true. But
while this model would be based on a speculative and empirically inaccessible
order of interactions of otherwise simultaneous measurements, the
updating models provided for sequential measurements are in tune with
the observed sequence of interactions of the consecutive measurements.
\end{enumerate}
\vspace{0.2in}
\noi \textbf{Acknowledgements.} This work has been supported by the
Friedrich Wilhelm Bessel Research Award of the Alexander von Humboldt
Foundation, the Hungarian National Research, Development and Innovation
Office, K-134275. I thank Ádan Cabello for valuable discussions. 

\section*{References}

\noindent \footnotesize

\begin{list}{--}{}
\item J. S. Bell, "On the problem of hidden variables
in quantum mechanics,'' \emph{Reviews of Modern Physics}, 38, 447-452 (1966) reprinted in J. S. Bell, \emph{Speakable and Unspeakable in Quantum Mechanics}, (Cambridge: Cambridge University Press, 2004).

\item A. Cabello, "Experimentally testable state-independent quantum
contextuality,'' \emph{Phys. Rev. Lett.}, 101, 210401 (2008).
\item A. Cabello, "Proposal for revealing quantum nonlocality via
local contextuality,'' \emph{Phys. Rev. Lett.}, 104, 220401 (2010).

\item A. Cabello, M. Gu, O. Gühne, and Z. Xu, "Optimal classical
simulation of state-independent quantum contextuality,'' \emph{Phys.
Rev. Lett.} 120, 130401 (2018).

\item O. Gühne, M. Kleinmann, A. Cabello, J. Larsson, G. Kirchmair,
F. Zähringer, R. Gerritsma, and C. F. Roos, \textit{\emph{"Compatibility
and noncontextuality for sequential measurements,'' }}\textit{Phys.
Rev. A, }\textit{\emph{81, 022121 (2010).}}

\item N. Harrigan, and R. Spekkens, "Einstein, incompleteness,
and the epistemic view of quantum states,'' \emph{Found. Phys.},
40, 125--157 (2010).

\item C. Held, "The Kochen-Specker theorem,'' \emph{The Stanford
Encyclopedia of Philosophy}, (2022) 

URL = https://plato.stanford.edu/entries/kochen-specker/\#question.

\item G. Hofer-Szabó, "Commutativity, comeasurability, and contextuality
in the Kochen-Specker arguments,'' \emph{Phil. Sci.}, 88, 483-510
(2021a).

\item G. Hofer-Szabó, "Three noncontextual hidden variable models
for the Peres-Mermin square,'' \emph{Eur. J. Phil. Sci.}, 11, 30
(2021b).

\item G. Hofer-Szabó, "Two concepts of noncontextuality in quantum
mechanics,'' \emph{Stud. Hist. Phil. Sci}, 93, 21-29 (2022).

\item G. Kirchmair, F. Zähringer, R. Gerritsma, M. Kleinmann, O.
Gühne, A. Cabello, R. Blatt and C. F. Roos, \textit{\emph{"State-independent
experimental test of quantum contextuality,'' }}\textit{Nature Letters}\textit{\emph{,
460, 494-498 (2009}}\emph{).}

\item M. Kleinmann, O. Gühne, J. Portillo, J. Larsson and A. Cabello,
"Memory cost of quantum contextuality,'' \emph{New J. Phys.} 13
113011 (2011).

\item B. Liu, X. Hu, J. Chen, Y. Huang, Y. Han, C. Li, G. Guo, and
A. Cabello, "Nonlocality from local contextuality,'' \emph{Phys.
Rev. Lett.}, 117, 220402 (2016).

\item B. R. La Cour, "Quantum contextuality in the Mermin-Peres
square: A hidden-variable perspective,'' \emph{Phys. Rev. A}, 79,
012102 (2009) .

\item B. R. La Cour, "Local hidden-variable model for a recent
experimental test of quantum nonlocality and local contextuality,''
\emph{Phys. Lett. A}, 381, 2230-2234 (2017).

\item D. Mermin, “Hidden variables and the two theorems of John Bell,”
\emph{Rev. Mod. Phys.}, 65 (3), 803-815 (1993).

\item A. Peres, “Incompatible results of quantum measurements,” \emph{Phys.
Lett. A}, 151, 107-108 (1990).

\item F. M. Leupold, M. Malinowski, C. Zhang, V. Negnevitsky, J.
Alonso, and J. P. Home, “Sustained state-independent quantum contextual
correlations from a single ion,” \emph{Phys. Rev. Lett. }120\emph{,}
180401 (2018).
\end{list}
\end{document}